# Fast spontaneous emission and high Förster resonance energy transfer rate in hybrid organic/inorganic plasmonic nanostructures

Hassnain Asgar, Liyan Jacob and Thang Ba Hoang

*Department of Physics and Materials Science*

*The University of Memphis, Memphis, TN 38152, USA*

Abstract: We report an experimental study of the plasmon-assisted spontaneous emission and the Förster resonance energy transfer between organic molecules and semiconductor colloidal quantum dots. The localized plasmonic field in the nanogap between a gold nano-popcorn's tips and a 5-nm separated gold film supports high photonic density of states and provides pathways for the light-matter interaction mechanisms. We demonstrate that, besides the total spontaneous emission rate enhancement factor up to 66 for quantum dots and molecules, the Förster resonance energy transfer efficiency and rate constant are simultaneously modified. While the energy transfer efficiency is reduced from 84% to 35% due to the non-radiative quenching effect and fast donor spontaneous emission rate, the energy transfer rate constant is significantly increased from 4 to 20 ns$^{-1}$. Our results have quantitatively elucidated decay mechanisms that are important toward understanding and controlling of the light-matter interaction at the nanoscale.

Recent developments in fabrication techniques have enabled the integration of nanomaterials (or emitters) with optical platforms, such as plasmonic nanocavities, at length scales much smaller than the diffraction limit.[1,2] A plasmonic nanocavity provides a highly concentrated electromagnetic field in a small volume where the high photonic density of states allow the spontaneous emission rate of an emitter to be significantly enhanced through the Purcell effect.[3,4] For instance, it has been shown that the intrinsic limits of emitters such as slow emission, low quantum yield or non-directional emission characteristics, can be significantly improved by integrating quantum emitters with optical nanocavities.[5-8] Integrating nanomaterials with optical nanocavities extends beyond an engineering challenge, as it offers rich physics of the light-matter interaction at the nanoscale which will eventually lead to exotic properties and applications.[5,9]

On the other hand, when two or more emitters are closely packed within a small mode volume generated by a plasmonic nanocavity, they can transfer energy from one to another through the non-radiative dipole-dipole interactions.[10-12] This process is called the Förster resonance energy transfer (FRET).[13] It has also been demonstrated also that under the influence of high photonic density of states provided by a plasmonic field, the FRET efficiency and rate are significantly modified.[14-17] In practice, the Purcell enhancement factor is determined as $F_P = \Gamma_{sp}^{cav}/\Gamma_{sp}^0 = \tau_{sp}^0/\tau_{sp}^{cav}$ where $\tau_{sp}^0, \tau_{sp}^{cav}$ are the intrinsic and reduced decay times of the emitters, respectively, while the FRET rate constant is determined as $\Gamma_{FRET} = \Gamma_{D_A} - \Gamma_D = 1/\tau_{D_A} - 1/\tau_D$ where $\tau_{D_A}, \tau_D$ are the decay times of donor with and without acceptor. In the presence of the plasmon nanocavity, because both $\tau_{D_A}, \tau_D$ are reduced and it is commonly a challenge to determine different decay mechanisms of the photo-excited carries. Therefore, it is important to quantitatively elucidate the contributions of different mechanisms, in a given plasmonic nanostructure, in order to understand and control the light-matter at the nanoscale.

In this work, we investigate the spontaneous emission rates together with the FRET between organic cyanine3 carboxylic acid (Cy3, donor) molecules and inorganic CdSe colloidal QDs (acceptor) under the influence of a plasmonic field. The interactions between these emitters with plasmonic gold (Au) nano-popcorns (NPs) were studied in various configurations. Unlike many other previously studied metallic nanoparticles with given symmetries such as nanospsheres[18,19] or nanocubes,[20,21] nano-popcorns[22,23] and nano-stars[24-26] exhibit somewhat non-uniform distribution of localized plasmonic fields over a set of spikes or tips. NPs and stars have recently attracted intensive attention in nanophotonics,[24,27] biological science[28,29] and materials science[30,31] thanks to their ability to localize electromagnetic fields in tiny hot spots at their highly engineered tips. Here, we demonstrated that by placing Au NPs on top of a Au film, separated by a thin (5 nm) dielectric polymer layer, both the spontaneous emission rates of Cy3 and QDs and their FRET rate constant are significantly enhanced. Our results indicate the highly intense electromagnetic field enhancement in the tiny gaps formed between NP's tips and the Au film can compensate for the loss due to non-radiative quenching and promotes the light-matter interaction at sub-10 nm scales.

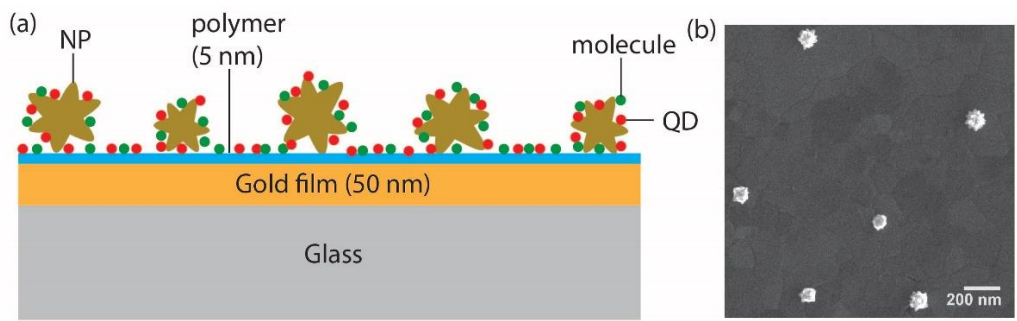

**Figure 1.** (a) Schematic of the sample design (b) SEM image of a completed sample.

The sample structure of this present study consists of colloidal synthesized Au NPs[24,25,27,32,33] with an approximate size of ~ 70 nm placed over a Au film (50 nm thick, fabricated by sputtering method) with a 5 nm polymer spacer layer sandwiched in between (Figure 1(a)). The polymer spacer layer was formed by alternative dip coating with polyelectrolytes (PE).[34] Specifically, five alternative positive PAH (poly(allylamine) hydrochloride) and negative PSS (polystyrenesulfonate) layers (1 nm each) were used. We observed that with a 5 nm spacer, the plasmon resonance could cover a broad spectral range from the absorption of the Cy3 dye to the emission of the QDs. The spikes (or tips) of the Au NPs were estimated, via Scanning Electron Microscopy (SEM) image (Figure 1(b)), to be less than 10 nm. Cy3 molecules (purchased from APExBIO) and CdSe QDs (Sigma Aldrich) solutions were diluted to 0.1 mM each and mixed together with Au NPs prior to the deposition to Au coated polymer film as schematically depicted in Figure 1a. The deposition technique was described elsewhere[34] and the surface coverage of NPs was less than 5%. In the experimental procedure, an ensemble of NPs, dressed with Cy3/QDs, were optically excited by an ultrafast laser (Coherent Chameleon Ultra II, 80 MHz, 150 fs). The excitation wavelength was chosen to be at 540 nm to overlap with the absorption spectrum of Cy3 (donors) (Figure 2a). The laser excitation spot was ~ 3 μm and average power was 5 μW. The photoluminescence (PL) signal was analyzed by a spectrometer (Horiba iHR550) and detected by a CCD (Charged-Coupled Device) camera. For the time-resolved single photon counting measurements, PL signal was filtered by the same spectrometer, guided through a second exit port and collected by a fast-timing avalanche photodiode. A time-correlated single photon counting module (PicoHarp 300) with a time bin of 4 ps was used to analyze the number of photons as a function of time when they arrived at the photodiode. Final lifetimes were obtained from fits to the data de-convolved with the instrument response function.[35] Absorption measurements were carried out using a Cary 100 UV-Vis spectrometer.

Figure 2(a) shows the absorption and emission spectra of both Cy3 dyes (dashed black) and CdSe QDs (dashed green) when they were dispersed on glass slides. The emission spectrum of Cy3 dye (blue) has a significant overlap with the absorption spectrum of the QD acceptor thereby allowing the FRET to occur. Further, the NPs' plasmonic resonance, as indicated by a broad reflection spectrum (shaded curve), plays role in a number of processes. First, it helps to increase the absorption rate of the donors. Second, this plasmonic resonance will increase the FRET rate through the interactions with both the donor and acceptor. Finally, due to the overlapping with both the emission spectra of Cy3 donor and QD acceptor (red), the plasmonic resonance will increase the spontaneous emissions rates of these emitters through the Purcell effect.

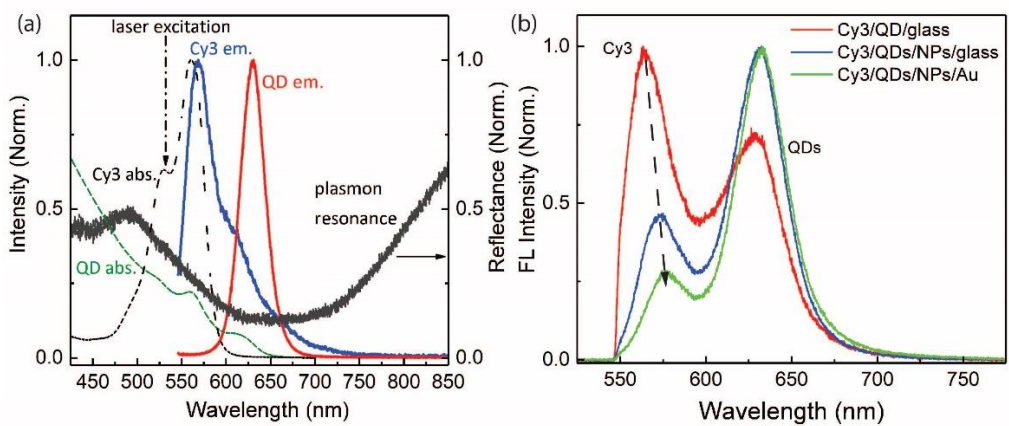

**Figure 2**. (a) PL emission spectra of Cy3 and QDs on glass slides, along with absorbance spectra of Cy3, QDs and NPs. (b) Emission spectra of combined Cy3/QDs with and without NPs on glass or Au film.

In Figure 2(b) the emission spectra for both Cy3 and QDs are shown for several configurations. For an easy comparison we show the normalized spectra to indicate the ratio between the emission peaks of Cy3 (at 570 nm) and QDs (at 630 nm). When the combined Cy3/QDs was on a glass slide (red) the emission spectrum showed a higher intensity for Cy3 dye. For a same dye/QDs concentration on a glass slide but with the presence of the NPs, the ratio between intensities of the Cy3 dye and QDs reduced, indicating that non-radiative decay has occurred.[5,36,37] The ratio of the peak intensities Cy3/QDs was further reduced when the glass slide was replaced by a Au film (separated by 5 nm from NPs), indicating that the Au film had an influence to the dynamics of the emitters. It is worth noting that while the QD emission peak remains fixed at around 630 nm, the emission peak of Cy3 dye shifted from 570 nm to 575 nm most likely due to the result of the spectral diffusion toward the maximum of the NP plasmonic resonance (~ 650 nm). In the sections below, we will point out that even though the Au film has a direct effect, namely the quenching effect of the emitters, the main contribution to the dynamics of the emitters is related to the tiny hot spots formed between the NPs' tips and the Au film.

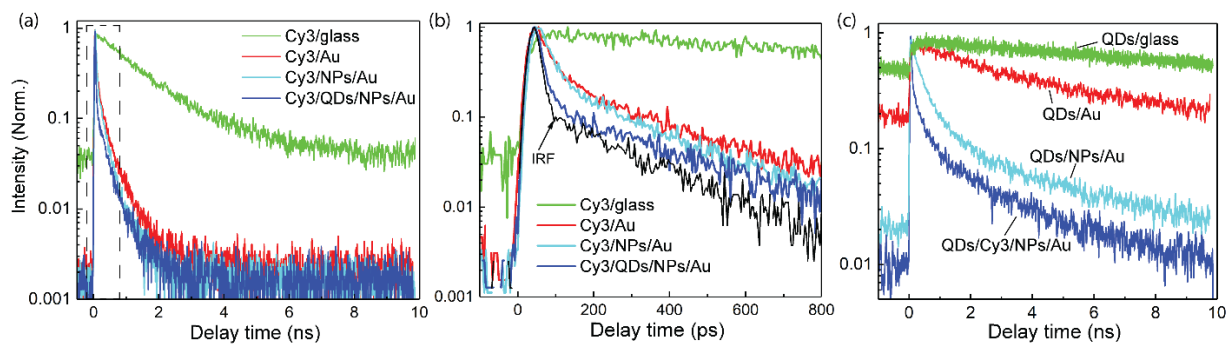

**Figure 3**. (a) Decay curves of Cy3 molecules measured at 570 nm for different conditions. (b) Same as (a) but plotted in a shorter time scale and (along with the system response function IRF) (c) Decay curves of the QDs measured at 630 nm for different configurations.

In order to gain further insight into the dynamics of the emissions and interactions between Cy3 dyes and QDs, we performed a series of experiments to determine the decay dynamics of photo-excited carries from these emitters. Figure 3(a) shows the decays curves of Cy3, measured at the maxima of the dye emission, in several configurations. Figure 3(b) shows the same data as presented in Figure 3(a) but on a shorter time scale (as indicated by dashed rectangle box) for easy viewing. First, as a control measurement, the green curve represents the intrinsic decay of the Cy3 molecules on a glass slide which results in a decay time of $1.20 \pm 0.01$ ns. The same dye concentration, when dispersed on a Au film, resulted in a much shorter decay time of $0.033 \pm 0.003$ ns (red) which is the direct consequence of the non-radiative decay due to the quenching effect.[5,37] Further, on a Au film, when NPs were included the decay time of the molecules further reduced to $0.028 \pm 0.002$ ns (Figure 3(b), cyan). To rule out a possible rate enhancement due to the NPs alone, we also performed the measurement of Cy3 molecules together with NPs on a glass slide (without Au film). The measured decay time, in this case, was $0.095 \pm 0.005$ ns (see *Table I* below) and was associated with a significant reduced PL signal intensity. These measurements indicate that the combination of NPs and Au film has resulted in an increase in the spontaneous emission rate of the Cy3 dye up to a factor of 43 (Purcell factor), which is directly related to the nanogap sandwiched in between a NP and the Au film. Finally, by including the QDs with Cy3 and NPs/Au film structure, the measured Cy3 decay time was $0.018 \pm 0.003$ ns (Figure 3(b), blue, limited by the instrument response function IRF) and was faster than its spontaneous emission decay time when the QDs were absence. This latter result is particularly interesting because it indicates that there was clearly an interaction between the dye and QDs, and roles of both NPs and the Au film. We also performed an additional control measurement where both Cy3 and QDs with NPs were on a glass slide (without Au film) and obtained a decay time of $0.077 \pm 0.003$ ns for Cy3 (see *Table I* below), which resulted from a combination of non-radiative quenching and FRET effect between Cy3 and the QDs.

In additional to the decay dynamics of Cy3 donor molecules in various configurations, we also investigated the dynamics of the QDs and the results are presented in Figure 3(c). On a glass slide, the intrinsic decay time of these QDs was measured to be $9.50 \pm 0.012$ ns (green, measured at 630 nm). When dispersed on a Au film, the decay time reduced to 3.035 ns (red) as a result of the direct quenching effect similar to the case of Cy3. When we included NPs with this latter case, i.e. QDs and NPs on a Au film, the decay time of the QDs significantly reduced to $0.145 \pm 0.004$ ns resulting in a spontaneous emission rate enhancement factor of 66 which is close to the value observed for Cy3 presented above. Further, we measured the QDs and NPs combination on a glass slide, the QD decay time was $0.163 \pm 0.004$ ns (*Table I*) and was associated with a significant reduction in the QD PL intensity (see Figure 4(b) below) due to the strong quenching effect by the NPs themselves. Finally, when we included all together (Cy3, QDs, NPs and Au film), the decay time of the QDs became even faster at $0.037 \pm 0.005$ ns. Comparing with the case where there were only QDs and NPs on Au film, the inclusion of Cy3 molecules has led to a further reduction of the QDs decay time (from 0.145 to 0.037 ns). The reduction of the QD acceptor decay time in this last configuration could be related to the modifications of the dipoles (both QDs and Cy3) in the presence of the NPs as well as change in the spectral overlap integral. In this current work, it is likely due to the latter as indicated by the fact that the emission peak of Cy3 donor was slightly shifted toward longer wavelength (Figure 2(b)) and modified the spectral overlap integral. It is also worth noting that, besides the fast decay component, the QDs decay dynamics in the presence of NPs and Au film also exhibited a weak and slower decay component (~ 2.5 ns) which is most likely associated with QDs that did not couple to the plasmonic resonance modes of the NPs (for instance, QDs in the area with no nearby NPs). This slow decay component has been reported before and we do not wish to examine again in this work.[6,7]

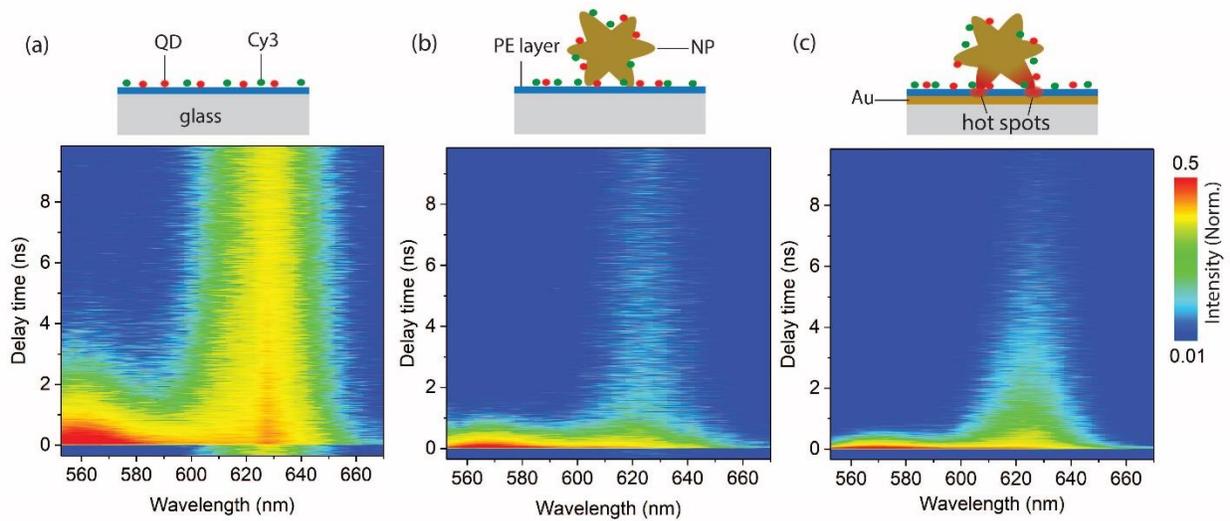

**Figure 4**. Decay dynamics of combined Cy3/QDs as functions of wavelengths in various configurations (a) on a glass slide (b) on a glass slide with NPs and (c) on a Au film with NPs. Insets on top show schematics of the sample structures for the measurements.

Results of the decay dynamics presented in Figures 3 were measured at the maximum emission wavelengths (~ 5 nm bandwidth) of Cy3 or QDs under different conditions. To have a better idea of the decay dynamics over the entire emission range of both the Cy3 and QDs, we scanned the grating of the spectrometer and measured the decay traces at different wavelengths with increments of 5 nm (which is comparable to the bandwidth transmitted by the spectrometer's exit slit width. These decay traces were then used to construct 2D false-color maps representing decay times vs. emission wavelengths. Figures 4 (a)-(c) displays such data for different Cy3-QD interaction conditions, which resulted in different decay dynamics. For Cy3/QDs on a glass slide (Figure 4(a)), we observed relatively long decay times for both of the emitters and their strong signal intensities. With the inclusion of the NPs, the decay times became faster and signal intensity was significantly reduced (Figure 4(b)), indicating the non-radiative quenching effect of the metallic NPs. Lastly, with Cy3/QDs and NPs on a Au film (Figure 4(c)), the decay time of Cy3 became even faster and was associated with relatively high signal intensity for QDs. This result indicates both the Purcell and FRET effects have occurred and is supported by the fact that if either only NPs or Au film were present, the signal was significantly quenched as described above. It is important to mention that due to the spectral overlap between donor and acceptor emissions (see Figure 1(a)), the emission from Cy3 in principle could directly contribute to the decay dynamics measured at 630 nm. However, due to the weaker emission from Cy3 molecules (as a result of the FRET effect, Figure 2(b)) and the fast decay component associated with them was not observed at the QD emission (Figures 4(b) – (c)), it is believed that there was a very small (or negligible) Cy3 contribution at this wavelength (630 nm).

Finally, we wish to quantitatively extract several important parameters related to various decay mechanisms. *Table I* summaries the decay times of Cy3 dye and QDs measured under different configurations. From these decay times, we can extract the FRET efficiency $E = 1 - \Gamma_D/\Gamma_{D_A}$ as well as the FRET rate constant $\Gamma_{FRET} = \Gamma_{D_A} - \Gamma_D$, where $\Gamma_{D_A}$ and $\Gamma_D$ are the decay rate of Cy3 with and without the QD presence, respectively. The results of both FRET efficiency and rate constants are shown in *Table II*.

*Table I. Decay components of Cy3 molecules and QDs in different configurations*

| Configuration | $\tau_{Cy3}$ (ns) (donor) | $\tau_{QD}$ (ns) (acceptor) | Main mechanism |
|---|---|---|---|
| Cy3/glass | 1.20 ± 0.01 | | Intrinsic |
| QDs/glass | | 9.50 ± 0.015 | Intrinsic |
| Cy3/QDs/glass | 0.193 ± 0.008 | 0.20 ± 0.008 | FRET |
| Cy3/NPs/glass | 0.095 ± 0.005 | | Quenching |
| QDs/NPs/glass | | 0.163 ± 0.004 | Quenching |
| Cy3/QDs/NPs/glass | 0.077 ± 0.003 | 0.152 ± 0.004 | Quenching/FRET |
| Cy3/Au | 0.033 ± 0.003 | | Quenching |
| QDs/Au | | 3.035 ± 0.011 | Quenching |
| Cy3/NPs/Au | 0.028 ± 0.003 | | Purcell |
| QDs/NPs/Au | | 0.145 ± 0.008 | Purcell |
| Cy3/QDs/NP/Au | 0.018 ± 0.002 | 0.037 ± 0.003 | Purcell/FRET |

*Table II. Extracted FRET efficiencies and rate constants from different configurations*

| | Cy3/QDs/glass | Cy3/QDs/NPs/glass | Cy3/QDs/NPs/Au |
|---|---|---|---|
| FRET efficiency | (83.9 ± 3.5) % | (18.9 ± 1.2) % | (35.7 ± 2.3) % |
| FRET rate constant | (4.3 ± 0.13) ns$^{-1}$ | (2.46 ± 0.12) ns$^{-1}$ | (19.8 ± 1.3) ns$^{-1}$ |

The extracted FRET efficiencies and rates of different configurations reveal the fact that the NPs have indeed reduced the FRET efficiency. On a glass slide, the efficiency reduced to 18.9% from 83.9% due to the non-radiative loss into metallic NPs. With the Au film included, the FRET efficiency was increased to 35% (*Table II*). The fact that the FRET efficiency was still lower compared with the intrinsic value, even with the presence of a Au film and NPs, suggested that the high spontaneous emission rate enhancement factors of Cy3 has limited the energy transfer efficiency between the two emitters. This is further confirmed by examining the appropriate FRET rate constants for each case. On a glass slide, the rate constant was 4.3 ns$^{-1}$ and this value was reduced to 2.46 ns$^{-1}$ with the inclusion of NPs, indicating loss due to the quenching effect. When both NPs and Au film were present, the rate constant increased to 19.8 ns$^{-1}$ which was larger than that on a glass slide. It can be understood that for this latter configuration the efficiency was lower than the intrinsic efficiency because the FRET rate constant 19.8 ns$^{-1}$ for this configuration was slower than the enhanced spontaneous emission rate of the Cy3 donor which was 36 ns$^{-1}$ ($\tau_{Cy3/NPs/Au}$ = 0.028 ns). Our results are consistent with the previous study by Sebastien et al. for a different system.[17] Finally, we like to emphasize that the contributions of the hot spots, formed between NPs' tips and the Au film, to the FRET efficiency and the increasing of the FRET rate constant were estimated to be in their lower bounds. This is because the mode volume generated by a hot spot was very small (consider the tip size < 10 nm) while the excitation spot was much larger (3 μm) and only molecules and QDs within the small area of the NPs tip(s) got enhanced. In our future work, by accurately determining the tip sizes, optimizing the gap thickness, and studying single NPs at different emitter concentrations, we expect a higher value for the FRET efficiency. Further, a computational study is needed to quantitatively verify the magnitude of the field enhancement in the hot spots.

In conclusion, we have investigated simultaneously the plasmon-assisted spontaneous emission and FRET of hybrid organics/inorganics at the length scale much smaller than the diffraction limit. We demonstrated that by integrating organic Cy3 donors and inorganic CdSe colloidal quantum dots with Au NPs and Au film, both high spontaneous emission enhancement and high FRET rate constant can simultaneously occur. Through various experimental configurations we conclude that the tiny hot spots formed between NPs' tips and closely spaced Au film played an important role. The high spontaneous emission rate of donor resulting from the hot spots in the nanogap has limited the FRET efficiency while allowing a high FRET rate constant. Our study thus provides an insight into the contributions and dynamics of various mechanisms that can be elucidated for the optimization of future high frequency and highly efficient nanophotonics devices.

This work was supported by the National Science Foundation (NSF) (Grant # DMR-1709612) and in part by a grant from The University of Memphis College of Arts and Sciences Research Grant Fund. This support does not necessarily imply endorsement by the University of research conclusions.